\RequirePackage{ifpdf}
\ifpdf 
\documentclass[pdftex]{sigma}
\else
\documentclass{sigma}
\fi

\begin{document}
\allowdisplaybreaks

\renewcommand{\PaperNumber}{017}

\FirstPageHeading

\ShortArticleName{Symmetry Methods in Plasma Physics}

\ArticleName{Applications of Symmetry Methods\\ to the Theory of
Plasma Physics}

\Author{Giampaolo CICOGNA~$^\dag$, Francesco CECCHERINI~$^\ddag$
and Francesco PEGORARO~$^\ddag$}

\AuthorNameForHeading{G. Cicogna, F. Ceccherini and F. Pegoraro}

\Address{$^\dag$~Dip. di Fisica and INFN, Largo B. Pontecorvo 3,
Ed. B-C, 56127 -- Pisa, Italy}
\EmailD{\href{mailto:cicogna@df.unipi.it}{cicogna@df.unipi.it}}

\Address{$^\ddag$~Dip. di Fisica, INFM and CNISM, Largo B.
Pontecorvo 3, Ed. B-C,  56127 -- Pisa, Italy}
\EmailD{\href{mailto:ceccherini@df.unipi.it}{ceccherini@df.unipi.it},
\href{mailto:pegoraro@df.unipi.it}{pegoraro@df.unipi.it}}

\ArticleDates{Received October 17, 2005, in f\/inal form January 20,
2006; Published online February 02, 2006}

\Abstract{The theory of plasma physics of\/fers a number of
nontrivial examples of partial dif\/ferential equations, which can
be successfully treated with symmetry methods. We propose three
dif\/ferent examples which may illustrate the reciprocal advantage
of this ``interaction'' between plasma physics and symmetry
techniques. The examples include, in particular, the complete
symmetry analysis of system of two PDE's, with the determination
of some conditional and partial symmetries, the construction of
group-invariant solutions, and the symmetry classif\/ication of a
nonlinear PDE.}


\Classification{35A30; 35Q60; 22E70}

\section{Introduction}

The theory of plasma physics of\/fers a number of highly nontrivial
examples of partial dif\/ferential equations, which can be
successfully treated with symmetry methods. The three dif\/ferent
examples we are going to consider may clearly emphasize the
reciprocal advantage of this ``interaction'' between plasma
physicists and symmetry theorists: for the latters, plasma theory
of\/fers the opportunity of testing their methods in concrete
problems; for the formers, symmetry techniques provide new
solutions to their equations.

Our f\/irst example actually deals with the Liouville equation. This
equation admits well known properties and a rich literature is
devoted to it; there are however various similar but not
equivalent forms of this equation, as we shall brief\/ly point out,
which admit  solutions with dif\/ferent properties, of course.
After some comments on the general peculiarities of our equation
and of its symmetry properties, we  shall provide some new,
physically relevant, solutions.

The second example deals  with a system of two coupled PDE's. This
system appears to be very rich in various and nontrivial symmetry
properties. We shall provide the  algebra of its exact Lie point
symmetries and examine some properties of this (inf\/inite
dimensional) algebra. The system will be  also compared with the
properties of similar equations coming from magnetohydrodynamics.
Next we shall study some invariant solutions  of this system, and
discuss their physical aspects.  This system also admits some
relevant  ``weak''   (conditional and partial) symmetries, as we
shall show, leading to  other families of solutions. In addition,
we shall discuss some peculiarities of the problem, considering,
in particular, the ef\/fect  of ``truncating'' the system (putting
some physical parameters equal to zero) on its symmetry
properties.

The f\/inal example is actually a ``symmetry classif\/ication''
problem. We will propose indeed a~nonlinear PDE containing two
arbitrary functions and examine how symmetry properties depend on
the choice of these functions, also with the help of the notion of
the equivalence group. The conclusion is that, apart from very
special cases,  only dilation exact symmetries are admitted by
some well def\/ined  classes of functions. In addition, one
nontrivial conditional symmetry is admitted, leading to a
particularly simple reduced ODE, which can be easily reduced to
quadratures.

In this paper we will be concerned only with Lie point symmetries,
see \cite{Ov,Ol,Ste,BK,Ga,Ib,BA}, under usual and standard
def\/initions and assumptions, as stated e.g.\ in \cite{Ol,BA}. For
a dif\/ferent approach to symmetries of dif\/ferential equations,
using dif\/ferential form methods, see the recent review
paper~\cite{Har}.

For all details concerning plasma physics, we refer to
\cite{JPA,PP} and to the papers quoted therein.

\section{Plasma kinetic equilibria described by Liouville equation}

Under suitable approximations and in appropriate experimental
conditions, the  conf\/iguration of an isothermal plasma embedded in
a stationary magnetic f\/ield  is 2-dimensional (planar: say, in the
plane $x$, $y$).   It  can be shown (see \cite{PP} and
ref.~therein) that such a conf\/iguration is completely  described
by a function $u=u(x,y)$, which is the component of the  magnetic
vector potential perpendicular to the $x$, $y$ plane, and which
satisf\/ies the classical ``elliptic'' Liouville equation
\begin{gather} \label{Li}
\nabla^2 u+\exp(2u) = 0 .
\end{gather}
Let us write, f\/irst of all, its most general 1-dimensional
solution $u=u_1(x)$ which depends  only on $x$:
\begin{gather} \label{L1d}
u_1(x) = - \ln\left(\frac{1} {|c|}\cosh (cx+c')\right), \qquad
c,c'={\rm const},
\end{gather}
with $c=1$, $c'=0$, this is called in plasma theory the
1-dimensional Harris sheet pinch solution.

As is well known (see e.g.\ \cite{a15,Ibl,Kis}), equation
(\ref{Li}) admits the inf\/inite dimensional  algebra of Lie point
symmetries generated by the vector f\/ields
\begin{gather} \label{XL}
X\ =\  \xi(x,y)\frac{\partial}{\partial x}
+\eta(x,y)\frac{\partial}{\partial y}
+\zeta(x,y)\frac{\partial}{\partial u},
\end{gather}
where $\xi(x,y)$, $\eta(x,y)$ are arbitrary harmonic conjugated
functions, and $\zeta(x,y)=-d\xi/dx$.

It can be remarked for the sake of completeness, that the
``hyperbolic" Liouville equation, analogous to \eqref{Li},
\[\frac{\partial^2u}{\partial x^2}-\frac{\partial^2u}{\partial
y^2}+\exp(2u)=0\] admits similar but not identical symmetry
properties: in particular, in the hyperbolic case the functions
$\xi$, $\eta$ in (\ref{XL}) can be arbitrary functions (see
\cite{FuS,PSa} and \cite[Vol.~1, p.~204]{Ib}).

Putting, in the case of equation~\eqref{Li},
\begin{gather}
\phi = \phi(z) = \xi+i\eta  \qquad {\rm with}\quad z=x+iy
\end{gather}
the function $\phi(z)$ is then an arbitrary holomorphic function
of the complex variable $z=x+iy$, and the symmetry property stated
above for equation \eqref{Li} implies that, choosing such
a~function~$\phi(z)$, one can construct continuous families of
solutions starting from any given solution to~(\ref{Li}).

On the other hand, it has been shown by Liouville
\cite{Kis,Liouville,Bate,Br} that the most general solution to
\eqref{Li} can be written in the form
\begin{gather} \label{Lga}
u(x,y)  = \ln\left(\frac{2|\gamma_z(z)|}{1+|\gamma(z)|^2}\right),
\qquad  z=x+iy,
\end{gather}
where the ``generating function'' $\gamma=\gamma(z)$ is another
arbitrary holomorphic function, and $\gamma_z(z)=d\gamma/dz$.  For
instance, the choice $\gamma=\exp (z)$  produces the above
mentioned $1$-dimensional Harris solution. The slightly more
general choice $\gamma(z)=k \exp(k\,z)+\kappa$, with
$\kappa=\sqrt{k^2-1}$, $k\in{\mathbb R}$, $|k|>1$,  gives the
solution
\[
u(x,y) = -\ln\big( \cosh(k\,x)+(\kappa/k) \cos(k\,y)\big)
\]
which describes a magnetic f\/ield conf\/iguration with a chain of
magnetic ``islands'' analogous to  ``cat's eyes" vortex chain in
hydrodynamics.

It can also be noted that, choosing $\gamma(z)=z^a$ $(a>0)$ in
\eqref{Lga}, one obtains radial solutions to~\eqref{Li}
\[
u = -\ln\left[{r\over {2a}}\left(r^a+{1\over
r^a}\right)\right],\qquad r^2=x^2+y^2
\]
which present a logarithmic singularity at the origin unless
$a=1$. More precisely, it can be shown the following
result~\cite{Br,cin}.

The most general solution to \eqref{Li} which is globally def\/ined
and satisf\/ies the ``normalization condition''
\[
\int_{{\mathbb R}^2}  \exp(2u)\, dx\, dy <  \infty
 \]
is the solution generated by $\gamma(z)=c(z-z_0)$, $c\in{\mathbb
R}$, $c\not=0$, i.e.\ a solution radially symmetric with respect
to a point $(x_0,y_0)\in{\mathbb R}^2$, of the form (with $k=1/c$)
\[ u(x,y) = \ln\left({2|k|\over{k^2+(x-x_0)^2+(y-y_0)^2}}\right)
 \]
which is known as the  Bennet pinch solution.

Putting $\gamma(z)=\exp\big(\beta(z)\big)$, one obtains from
\eqref{Lga} this other useful expression of the general solution
to~\eqref{Li}
\[
u(x,y)  = -\ln\left(\frac{\cosh({\rm
Re}\beta(z))}{|\beta_z(z)|}\right) .
\]
This formula and the above equation \eqref{Lga} are the two
expressions commonly used in the applications in plasma physics.
It can be noted, incidentally, that if one considers the other
equation\begin{gather} \label{min} \nabla^2u-\exp(2u)\ =\
0\end{gather} \big(or $\nabla^2u+\exp(-2u)=0$\big) slightly
dif\/ferent from \eqref{Li}, several dif\/ferent representations are
known of its general solutions \cite{a15,Kis}, but not all admit
an analogous expression for our equation~\eqref{Li}. For instance,
equation~\eqref{min} admits the two dif\/ferent solutions depending
only on $x$
\[
u^{(1)}(x) = -  \ln\big|\sinh  x\big| \qquad{\rm and} \qquad
u^{(2)}(x)\ =\ - \ln\big|\sin  x\big|
\]
whereas the solution given in \eqref{L1d} is {\it the most
general} solution to \eqref{Li} depending on~$x$. One can also
show that the elliptic Liouville equation in the form~\eqref{min}
does not possess solutions valid in the entire plane, while
equation \eqref{Li} does possess such solutions.

For a full discussion about this (not marginal) point and a
comparison between the possible expressions representing the
various solutions of these Liouville equations, see~\cite{Cro}. It
can be also noted that equation~\eqref{min} has the same symmetry
properties as equation~\eqref{Li}.

Another peculiarity shared by all these types of Liouville
equations is the possibility of stu\-dying and f\/inding their
solutions by means of B\"acklund transformations. Also in this
approach, elliptic and hyperbolic equations actually behave a
little bit dif\/ferently: see e.g.~\cite{Kis,AnIb,KCK}. We do not
insist on this topic, which goes beyond the scope of the present
paper.

Coming back to our equation  \eqref{Li}, let us consider for
instance the Lie symmetry determined by $\phi(z)=z^2$, i.e.
\[ X = \left(x^2-y^2\right) \frac{\partial}{\partial
  x}+2xy\frac{\partial}{\partial  y}-2x\frac{\partial}{\partial u} .
\]
Starting from the Harris solution, we then obtain the  following
family of solutions, describing a~continuously deformed family of
curved plasma sheet conf\/igurations
\[
u(x,y) = -\ln\left[\left(1-2\lambda x+\lambda^2r^2\right)\cosh
\frac{x-\lambda r^2}{1-2\lambda x+\lambda^2r^2}\right],
 \]
where $r^2=x^2+y^2$, $\lambda\in {\mathbb R}$ is the group
parameter. The generating function $\gamma$ is now
\[
 \gamma=\gamma (z,\lambda)=\exp\left({z\over{1-\lambda z}}\right);
\]
in this way, the essential singularity of the generating function
of Harris solution is shifted from inf\/inity to $x=1/\lambda$.

Even more in general, one can perform an arbitrary holomorphic
transformation
\[ z\to\widetilde z=\psi(z) \]
and consider then the new generating function $\widetilde
\gamma(z)\equiv \gamma\big(\psi(z)\big)$; thanks to (\ref{Lga}),
one obtains in this way other solutions to (\ref{Li}). Actually,
it can be shown \cite{a15} that any solution  to (\ref{Li}) can be
obtained in this way;  this means in particular that any solution
can be transformed  (locally) into the 1-dimensional solution
(\ref{L1d}) by means of a suitable holomorphic transformation
$z\to \psi(z)$.

The close relationship between the generating function $\gamma(z)$
of any solution  of \eqref{Li} and symmetry properties, can be
also emphasized by the following result \cite{PP}.
\begin{proposition}
Let $u=u(x,y)$ be any solution  to \eqref{Li}, $\gamma(z)$ its
generating function and  $X_0$ the~Lie vector field which leaves
invariant  this solution: then
\begin{gather} \label{gf}
\gamma_z(z)\phi_0(z) = i\gamma(z),
\end{gather}
where  $\phi_0(z)$ is the holomorphic function which determines
$X_0$.
\end{proposition}

For instance, for the  above solution  (\ref{L1d}), one has
$\gamma=\exp(cz+c')$ (with $ c,c'\in {\mathbb R}$), and then from
(\ref{gf}) $\phi_0=i/c$, i.e.~$X_0=\partial/\partial y$, as
expected.

To conclude this section, let us point out, among many other
solutions, the three following examples, for their special
interest to plasma physics (see~\cite{PP}).

Choosing as  generating function
\[
\gamma(z)=(1+pz)^{1/p},  \qquad p>0
 \]
one obtains a   one-parameter family of  magnetic conf\/igurations
with circular f\/ield lines, which includes the  Bennet pinch for
$p=1$, ring-like structures for $0< p <1$, and which gives the
Harris solution  at the limit $p\to 0$. Instead, the solution
generated  by
\[ \gamma(z)=p\, {\rm Erf}(z) \]
produces a sort of bar-like structure in the plasma current
distribution. Finally, starting from
\[
\gamma(z)={z+p\over{z-p}}\exp z
\]
one gets a solution  which can be used to model some important
features of the  Earth magnetotail.

\section{A system of PDE's for magnetized plasmas}

The case we are now considering, which is the most elaborate and
richest in dif\/ferent symmetry  properties, is a system of two
PDE's for the two functions  $u=u(x,y,t)$, $v=v(x,y,t)$.

The system is the following
\begin{gather*}
\frac{\partial}{\partial t}\big(u-\nabla^2 u\big) -
\big\{u-\nabla^2 u,v\big\} + \big\{u,\nabla^2 v \big\} = 0,
\\
\frac{\partial} {\partial t} \big(\nabla^2 v\big) -
\big\{u,\nabla^2 u\big\}+\big\{v, \nabla^2 v \big\} = 0,
\end{gather*}
where the $\{\cdot,\cdot\}$ is def\/ined by
\[
\{f,g\}\ = {\partial f\over{\partial x}}{\partial g\over{\partial
y}}-{\partial g\over{\partial x}}{\partial f\over{\partial y}}, \]
or, written in a more ``symmetric'' and compact form,
\begin{gather} \label{EU}
{\partial\over{\partial t}}\big(u-\nabla^2 u \pm \nabla^2 v\big)+
\big\{v\pm u,u-\nabla^2 u \pm \nabla^2 v \big\}  = 0   .
\end{gather}
In this system of PDE's the  functions $u(x,y,t)$, $v(x,y,t)$
describe respectively the time dependent magnetic and electric
potentials of a 2-dimensional planar conf\/iguration of a  f\/luid
plasma,  embedded  in a strong magnetic f\/ield  orthogonal to the
$x$, $y$ plane and with a shear magnetic  f\/ield in this plane.

\subsection{The Lie algebra of symmetries}

With the help of some appropriate computer package, e.g.\
\cite{CHW,Her,Ba}, it is possible to show the following result:
\begin{proposition}
The system \eqref{EU} admits the Lie algebra of symmetries
generated by the following operators:
\begin{gather} \label{X123}
X_1 = {\partial\over{\partial t}}, \qquad X_2 =
y{\partial\over{\partial x}}-x{\partial\over{\partial y}}, \qquad
X_3 = \frac{\partial}{\partial u} ,
\\
\label{X2} X_4 = -t y {\partial\over {\partial x}}+t x
{\partial\over{\partial y}}+ {x^2+y^2\over {2}} {\partial
\over{\partial v}}
\end{gather}
and by the family of operators, depending on three arbitrary
smooth functions $H(t)$, $A(t)$, $B(t)$:
\begin{gather}  X_H = H(t)\frac{\partial}{\partial v},   \\
 \label{X1}
X_{(A,B)}= A(t)  {\partial\over{\partial x}}+ B(t)
{\partial\over{\partial y}}+\Big(x {B_t}  - y {A_t}
\Big){\partial\over{\partial v}},
\end{gather}
where $A_t=dA/dt$, etc., which generate an infinite dimensional
subalgebra (actually: an ideal).
\end{proposition}

The nonvanishing commutation rules between the above def\/ined
symmetry generators are the following:
\begin{gather*}
[X_1 , X_4] = -X_4, \qquad [X_1 , X_H] = X_{H_t},\qquad
[X_1  , X_{(A,B)}]  =  X_{(A_t,B_t)},\\
[X_2,  X_{(A,B)}] = X_{(-B,A)},\qquad [ X_4, X_{(A,B)}] =
-X_{(-tB,tA)}, \qquad  [X_{(A,B)} ,  X_{(C,D)}]   =  X_{\widetilde
H},
\end{gather*}
\noindent where   $\widetilde H = AC_t - BD_t -C B_t + DA_t$.

The operators $X_1$, $X_2$, $X_3$ generate respectively time
translation, space rotation and the translation $u\to u+k$, and
together with the operator $X_4$ form a subalgebra of the algebra
of the symmetries; the operators $X_{(A,B)}$ include in particular
space translations $\partial /\partial x$ and $\partial /\partial
y$. The operator $X_{H}$ generates the translation $v\to v+H(t)$;
this operator and $X_3$ (which is the center of the algebra)
  do not change either the magnetic f\/ield
or the plasma velocity. Some properties of the two remaining
operators $X_4$ and $X_{(A,B)}$ will be considered in the next
subsection.

\subsection{Symmetry properties and group-invariant solutions}

Before considering  group-invariant solutions of system
\eqref{EU}, let us  look for an optimal system (see
e.g.~\cite{Ol}) of $1$-dimensional subalgebras of the algebra of
symmetries listed in Proposition~2. It is easily seen that in the
adjoint representation the action of the operators $X_H$ and
$X_{(A,B)}$ on all the vector f\/ields $X_\alpha$ (where $X_\alpha$
denotes any of the operators def\/ined above) has the ef\/fect of
transforming the functions $H(t)$, $A(t)$, $B(t)$ and does not
touch the other operators; the same happens for the action of any
$X_\alpha$ on $X_H$ and $X_{(A,B)}$. For instance, one has:
\begin{gather*}
 {\rm Ad}\big(\exp(\lambda X_{(A,B)})\big) {\partial\over{\partial t}}
= {\partial\over{\partial t}}+\lambda
X_{(A_t,B_t)}-{\lambda^2\over
2}(AA_{tt}-BB_{tt}){\partial\over{\partial v}},\\
 {\rm Ad}\big(\exp(\lambda X_{(A,B)})\big) X_{(C,D)} =
X_{(A,B)}-\lambda(AC_t-BD_t-CB_t+DA_t) {\partial\over{\partial v}}
.
\end{gather*}
In view of the physical applications,  it is more convenient to
consider the functions $H$, $A$, $B$ as {\it arbitrary} functions,
and therefore we do not include in the classif\/ication the
operators $X_H$, $X_{(A,B)}$. Restricting to the operators $X_1$,
$X_2$ and $X_4$ (notice that $X_3$ acts trivially on all
the~$X_\alpha$), an optimal system is
\[X_1 =  {\partial\over{\partial t}} ,\qquad
X_2 =  y{\partial\over{\partial x}}-x{\partial\over{\partial y}},
\qquad X_a = a {\partial\over{\partial t}}+X_4,
\]
where $a$ is a real parameter.

The equations for time-invariant or rotationally-invariant
solutions can be easily deduced; on the other hand, they are
actually very special solutions, of no  general interest. The same
is true for solutions invariant under $X_H$ (i.e.\ solutions with
$v\equiv 0$). We then consider in some detail the symmetries $X_4$
and $X_{(A,B)}$. It is convenient to examine f\/irst the symmetry
$X_{(A,B)}$.

1)  The symmetry $X_{(A,B)}$ expresses the property that, if
$u(x,y,t)$, $v(x,y,t)$ is a solution of~\eqref{EU}, then
also\footnote{The parameter $\lambda$  which should be introduced
to parametrize this family of solutions can be clearly absorbed in
the (arbitrary) functions $A$ and $B$.}
\begin{gather*}
 \widetilde u(x,y,t):=u(x-A(t),y-B(t),t)  , \\
 \widetilde v(x,y,t):=xB_t-yA_t-{1\over
2}(AB_t-A_tB)+v\big(x-A(t),y-B(t),t\big)
\end{gather*}
solve our system~\eqref{EU}, for any $A(t)$, $B(t)$. It must be
noticed that this corresponds to a change of spatial coordinates
into a moving frame which produces in turns the additional term
$xB_t-yA_t-(1/2)(AB_t-A_tB)$ in the component $v$.   Since $v$ is
proportional to the electric potential, we can interpret this
symmetry as expressing the fact that  a time dependent, spatially
uniform, electric f\/ield imposed on the system induces a uniform
time-dependent electric drift.

We want now to look for the solutions to~\eqref{EU} which are {\it
invariant} under $X_{(A,B)}$. Let us recall that if a system of
PDE's  for the $q$ functions $u_a=u_a(x)$ of the $p$ variables
$x_i$ admits a symmetry $X = \xi_i \partial/\partial x_i+\zeta_a
\partial/\partial u_a$,
 $i=1,\ldots, p$, $a=1,\ldots , q$,
then the $X$-invariant solutions must satisfy the $q$ equations
\begin{gather}\label{ic}
 \zeta_a-\xi_i{\partial u_a\over{\partial x_i}} = 0 .
\end{gather}
These equations  in our case  take  the form
\begin{gather} \label{inv1}
A u_x+B u_y=0, \qquad A v_x+B v_y=xB_t-yA_t.
\end{gather}
The f\/irst equation  can be interpreted as the requirement  that
the displacement produced by the electric drift is parallel to the
f\/ield lines of the shear magnetic f\/ield, while the second
expresses the requirement   that the electric potential of the
displaced plasma element remains  constant.

Instead of looking directly for the invariant solutions to
\eqref{EU}  starting from (\ref{inv1}), it can be more interesting
to give preliminarily the expression of equations \eqref{EU} when
written in the ``canonical coordinates'' (or ``symmetry-adapted
coordinates'') \cite{Ol,BA}. In our case   these coordinates can
be chosen as follows
\[ s=B(t)x-A(t)y , \qquad  w=\frac{A(t)x+B(t)y} {A^2(t) +B^2(t)}, \]
where $s$ is (together with $t$) an invariant coordinate under the
symmetry $X_{(A,B)}$, and $w$ is the coordinate ``along the
action'' of the symmetry, i.e.\ $X_{(A,B)} w=1$. Notice also that
$\{f,g\}_{x,y}=\{f,g\}_{s,w}$.

Putting, as indicated by (\ref{inv1}),
\begin{gather} \label{VW}
  u=U(s,w,t), \qquad  v=Q(x,y,t)+V(s,w,t)
\end{gather}
where $Q$ is a  quadrupolar term  given by
\[ Q= {1/2\over{A^2+B^2}}\big((A_tB+AB_t)\big(x^2-y^2\big)-2xy(AA_t-BB_t)\big),
\]
we obtain from \eqref{EU} the following system
\begin{gather}
{\partial\over{\partial t}}\big(U-\widetilde \nabla^2 U \pm
\widetilde \nabla^2 V\big)+ \big\{V\pm U,U-\widetilde \nabla^2 U
\pm
\widetilde \nabla^2 V \big\} \nonumber\\
\qquad{}+ {2s\over{(A^2+B^2)^2}} {\partial\over\partial
w}\big(U-\widetilde \nabla^2 U\pm \widetilde \nabla^2 V\big) =
0,\label{Cor}
\end{gather}
where $\widetilde \nabla^2=(A^2+B^2){\partial^2/\partial s^2}+
(A^2+B^2)^{-1}{\partial^2/\partial w^2}$. Here the dif\/ferent role
of the two variables~$s$ and $w$ is  evident: in particular -- as
expected -- there is no explicit dependence on $w$, whereas the
last term, proportional to $s$, can be interpreted as a
Coriolis-type contribution.

Now,  if one looks for invariant solutions, i.e.\ for solutions of
the form
\[
u=U_0(s,t) ,\qquad v=Q(x,y,t)+V_0(s,t)
 \]
the above equations (\ref{Cor}) become
\[
{\partial\over{\partial
t}}\big(U_0-\big(A^2+B^2\big)U_{0,ss}\big)=0, \qquad
{\partial\over {\partial t}}\big(\big(A^2+B^2\big)V_{0,ss}\big) =
0 .
\]
These clearly imply
\[ U_0-\big(A^2+B^2\big)U_{0,ss} = F(s) ,\qquad
\big(A^2+B^2\big)V_{0,ss} = G(s),
 \]
where $F(s)$, $G(s)$  are arbitrary functions. Notice   that
${\partial^2/\partial x^2}+{\partial^2/\partial y^2}=(A^2+B^2)
{\partial^2/\partial s^2}$ and that these equations are actually
ODE's, indeed the variable $t$ here appears  merely as
a~parameter.

Special physically relevant solutions of this reduced system (and
of the system  \eqref{EU}, of course) can be easily obtained
starting from particular choices of the arbitrary functions
$A(t)$, $B(t)$, $F(s)$, $G(s)$ introduced above (see~\cite{JPA}
for some explicit examples).

2) We now consider the symmetry  $X_4$   of our system \eqref{EU}:
it implies that  if $u(x,y,t)$, $v(x,y,t)$ is a solution, then
also
\begin{gather*}
\widetilde u(x,y,t):=u\big(x \cos(\lambda t)+y \sin(\lambda t),
-x\sin(\lambda t)+y \cos(\lambda t),t\big),\\
\widetilde v(x,y,t):=v\big(x \cos(\lambda t)+y \sin(\lambda t),
-x\sin(\lambda t)+y \cos(\lambda
t),t\big)+\lambda{x^2+y^2\over{2}}
\end{gather*}
is a family of solutions to \eqref{EU} for any $\lambda\in{\mathbb
R}$. This represents a sort of rotated solutions with angular
velocity $\lambda$ plus a radial term in the component~$v$ which
gives the additional velocity f\/ield corresponding to the rotation.

The solutions to \eqref{EU}, which are   invariant  under
symmetry (\ref{X2}), must satisfy  the invariance condition
\[ y{\partial
u\over{\partial x}}-x{\partial u\over{\partial y}} = 0, \qquad
y{\partial v\over{\partial x}}-x{\partial v\over{\partial y}} =
-{r^2\over {2t}},
 \]
where $r^2=x^2+y^2$. It is easy to f\/ind that they are of the form,
with $ \theta=\arccos(x/r) $,
\begin{gather}\label{cut}
 u=U^{(0)}(r,t),\qquad
v={r^2 \over{2t}} \theta +V^{(0)}(r,t),
\end{gather}
where $U^{(0)}(r,t)$, $V^{(0)}(r,t)$ satisfy  the  {\em linear
and uncoupled} homogeneous PDE's
\begin{gather*}
r^2U^{(0)}_{rrr}-2rt  U^{(0)}_{rrt}+r  U^{(0)}_{rr}-2tU^{(0)}_{rt}
-r^2U^{(0)}_r+2rtU^{(0)}_t+3U^{(0)}_r=0,  \\
2rtV^{(0)}_{rrt}-r^2V^{(0)}_{rrr}+2tV^{(0)}_{rt}-rV^{(0)}_{rr}+5V^{(0)}_r=0
.
\end{gather*}
A simple solution  of the f\/irst equation is, e.g.,
$U^{(0)}=r^2t$, whereas the second one admits solutions of the
form $V^{(0)}=r^a t^b$  with $b=(a^2-2a-4)/2a$, for any $a$.   It
can be noted in particular that the special form of the component
$v$ in \eqref{cut}, which has a f\/ixed term containing a~``cut''
discontinuity, looks as a spiral and expresses the fact that this
solution  contains a $\theta$-independent azimuthal electric
f\/ield, vanishing for $t\to\pm\infty$. Solutions of this form are
not new in plasma theory and in magnetohydrodynamics: see
e.g.~\cite{Sam} and \cite[Vol.~1, p.~393]{Ib}. See also
Subsection~3.4 for other  remarks and comments on various related
physical aspects.

It is remarkable that, in both cases of symmetries (\ref{X1}) and
(\ref{X2}), the invariant solution  $u$ satisf\/ies a  linear
equation  (therefore linear superposition principle holds), and
that the same is true for $v$, apart from a f\/ixed additional term
(for some comment on this point we refer to \cite{JPA}).

3) Finally, one can also look for solutions invariant under
symmetries of the form, e.g.,
\begin{gather} \label{THA}
  {\partial\over{\partial t}}+X_H ,\qquad {\partial\over{\partial
t}}+X_{(A,B)}
\end{gather}
(or similar combinations: see our remark  on the optimal system of
symmetries at the beginning of this section). Clearly, the most
convenient choice of symmetry to be considered, as well as the
choice of the arbitrary functions $H$, $A$, $B$ involved, can be
suggested by the specif\/ic case to be examined or by the
experimental conf\/iguration. Just to give some examples, let us
consider the two symmetries \eqref{THA}.

For the f\/irst symmetry, it can be easily seen that the invariant
solutions have the form
\[ u\ =\ U(x,y),\qquad v=T(t)+V(x,y),\]
where $T_t=H(t)$ and $U$, $V$ satisfy the system
\[
\big\{U-\nabla^2U,V\big\}=\big\{U,\nabla^2V\big\}, \qquad
\big\{U,\nabla^2U\big\}=\big\{V,\nabla^2V\big\} .
\]
Considering instead the symmetry $\partial/\partial t+X_{(A,B)}$,
one f\/inds that the invariant solutions have the form
\begin{gather*} u=U\big(x-\alpha(t),y-\beta(t)\big),\\
v=\big(x-\alpha(t)\big)B-\big(y-\beta(t)\big)A +\alpha B-\beta
A+V\big((x-\alpha(t),y-\beta(t)\big),
\end{gather*}
where $\alpha_t=A$, $\beta_t=B$, and $U$, $V$ must satisfy a
system of nonlinear PDE's.

\subsection{``Weak'' symmetries}

The above system \eqref{EU} admits also several interesting
``non-exact'' (here generically called  ``weak'') symmetries; we
are going to consider an example of conditional symmetry (see
\cite{BC, BC1,FT,LW,W2,FK}), and one of partial symmetry
(according to the def\/inition given in \cite{CG}, see also
\cite{CC} and below).

As well known, any vector f\/ield $X = \xi_i \partial/\partial
x_i+\zeta_a \partial/\partial u_a$ is a~conditional symmetry for
a~dif\/ferential equation (or a system thereof) $\Delta=0$ if the
system $\Delta=0$ enlarged with the invariance condition
\eqref{ic} admits some solution. If this is the case, a
conditional symmetry allows the reduction of  the initial equation
into a reduced form, and in this way one can obtain  other
invariant solutions: see  \cite{BC, BC1,FT,LW,W2,FK}. See also
\cite{OR,Ol3,Pu, PS1,Odm,ZT,Pop,ZTP,IY} for   careful discussions
about various related problems and reduction procedures.

An  example of conditional symmetry leading to interesting
solutions to \eqref{EU} is the following (more precisely, it is a
``contact conditional symmetry''):
\[
X = u_y{\partial\over{\partial u}}+v_x{\partial\over{\partial v}}.
\]
 From this symmetry, one f\/inds e.g.\ solutions to \eqref{EU} of this form
\[
 u=\sin\big[k(x-T(t))\big], \qquad v=\exp(\pm \kappa y)-T_t(t) y ,
\]
where $k,\kappa\in{\mathbb R}$ with $k^2-\kappa^2=1$ and $T(t)$ is
arbitrary, or similar solutions where the functions $\sin$ and
$\exp$ are exchanged. Solutions of this form are interesting due
to the presence of terms depending on $x-T(t)$ describing
generalized wave propagation.

It is known that there are some delicate points related to the
def\/inition of conditional symmet\-ries: see \cite{OR,Ol3,Pu,
PS1,Odm,ZT,Pop,ZTP}; this is actually related to the introduction
of a~subtler classif\/ication of the notion of conditional symmetry
\cite{Cw,CL}. We do not deal here with this problem, and we prefer
to consider a particularly interesting example of partial symmetry
for our system \eqref{EU}.

While exact symmetries transform any solution  of the given
problem into  another solution, conditional symmetries do not
transform -- in general -- solutions into solutions, as well
known. Partial symmetries play in a sense an intermediate role,
indeed they transform solutions belonging to a proper subset of
solutions into solutions in the same subset. This subset is
def\/ined in the following way. Let us assume  that some vector
f\/ield $X$ is {\it not} an exact symmetry for a~dif\/ferential
equation (or a system) $\Delta=0$: this means that, denoting by
$X^*$ the appropriate prolongation of $X$, one has
$X^*(\Delta)|_{\Delta=0}\not= 0$. Let us then introduce the
condition
\begin{gather}
\Delta^{(1)}:=X^*(\Delta) =  0\label{D1}
\end{gather}
as a new equation, and consider the enlarged system
\begin{gather}
\Delta=\Delta^{(1)} = 0\label{DD} .
\end{gather}
It is clear that if $X$ is an exact symmetry of this enlarged
system, then the subset ${{\cal S}}^{(1)}$ of the simultaneous
solutions of this system (if not empty, of course) is a
``symmetric set of solutions''  to $\Delta=0$, i.e.\ a proper
subset of solutions which have the property of being mapped   the
one into another by the vector f\/ield $X$. It is also clear that
this property is not shared by the other  solutions to $\Delta=0$
not belonging to the subset ${{\cal S}}^{(1)}$.   In  principle,
this procedure can be iterated, see \cite{CG,CC}, but no iteration
is necessary in the  example we are considering.

The vector f\/ield we want to deal with is
\begin{gather}
X  =  u {\partial\over{\partial u}}\label{pp} .
\end{gather}
First of all, notice that this is  trivially a conditional
symmetry for \eqref{EU}, indeed the invariance condition simply
amounts in this case to look for the special solutions to
\eqref{EU} with $u=0$, i.e.\ to the hydrodynamic limit. More
interestingly, we now show that the vector f\/ield (\ref{pp}) is
 a~nontrivial partial symmetry: notice that considering this vector
f\/ield as a partial symmetry corresponds to looking for solutions
to \eqref{EU} with the property that the component $u$ admits a
scaling property, i.e.\  for solutions $u$, $v$ such that also
$\lambda u$, $v$ solve \eqref{EU} for  $\lambda\in{\mathbb R}$.
Applying the prolongation $X^*$ to the system \eqref{EU}, and
combining the resulting equation $\Delta^{(1)}=X^*(\Delta)$ with
\eqref{EU}, one gets the new condition
\begin{gather}
\big\{u,\nabla^2 u\big\}=0, \label{pdp}
\end{gather}
which is then the condition characterizing the subset ${{\cal
S}}^{(1)}$ of solutions with the above specif\/ied property. It is
easy to verify that the system of the three equations \eqref{EU}
and (\ref{pdp}) is symmetric under (\ref{pp}), showing that
(\ref{pp}) is indeed a partial symmetry for \eqref{EU}. Thanks to
\eqref{pdp}, we deduce this equation for $v$
\[  {\partial\over{\partial t}}\big(\nabla^2 v\big)+\big\{v,\nabla^2 v\big\} = 0 \]
which does not contain $u$,   and is equivalent to the
two-dimensional Euler equation for an incompressible f\/luid. On the
other hand, the  equation  for $u$ is  linear in $u$: obviously,
this agrees with the presence of the (partial) symmetry given by
(\ref{pp}).

Before considering examples of particular simple  solutions, let
us come back to an important physical feature of  our initial
system of dif\/ferential equations \eqref{EU}. This system indeed
would actually contain  some physical parameters  that we have
normalized to the unity up to now. There are physical situations,
however, where one of these parameters turns out to be  negligible
and can be put equal to zero with a good approximation \cite{JPA}:
this  coef\/f\/icient multiplies the term $\{u,\nabla^2 v\}$   in
\eqref{EU}. For this reason, it can be interesting to repeat
calculations determining symmetry properties and solutions of the
approximate system. The quite surprising result is summarized in
the following
\begin{proposition}
The truncated system
\begin{gather} \label{a1}
{\partial\over{\partial t}}\big(u-\nabla^2 u\big)-\big\{u-\nabla^2
u,v\big\} =0 , \qquad
     {\partial\over{\partial t}}\big(\nabla^2 v\big)
  -\big\{u,\nabla^2 u\big\}+\big\{v,\nabla^2 v\big\} = 0
\end{gather}
admits precisely the same exact symmetries as the original one
\eqref{EU}. The same is also true if one or both of the other
equations
\begin{gather} \label{a2}  \big\{u,\nabla^2 v\big\}=0, \qquad
\big\{u,\nabla^2 u\big\}=0
\end{gather}
are appended to the above system (therefore, even if the partial
symmetry \eqref{pp} is taken in consideration also within this
approximation).
\end{proposition}

Now observing that condition  (\ref{pdp}) implies $\nabla^2
u=A(u,t)$  for some smooth function $A$, and similarly $\{
u,\nabla^2 v\}=0$ implies $\nabla^2 v=B(u,t)$, the   system
(\ref{a1})--(\ref{a2}) becomes
\begin{gather*} (1-A_u)\big(u_t-\{u,v\}\big)=A_t  ,\qquad
\nabla^2 u=A(u,t) , \\
   -B_u\big(u_t-\{u,v\}\big)=B_t  , \qquad  \nabla^2v=B(u,t) .
\end{gather*}
Therefore, assuming e.g.\ $A_u=1$ forces $A_t=0$; instead if
$A_u\not=1$, $B_u\not=0$, but $A_t=B_t=0$, the above system takes
the very simple form
\begin{gather}
u_t=\{u,v\},\qquad \nabla^2 u=A(u), \qquad \nabla^2v=B(u).
\label{AB}
\end{gather}
Elementary solutions of  this system (which are  also solutions to
\eqref{EU}, of course) can be imme\-diate\-ly found, for instance,
just to give some simple examples,
\[ u=c_1\sin(k(x-t))+c_2\sin(k(y-t))+c_3), \qquad v=x-y, \]
where $c_i$, $k$ are arbitrary constants, or also $u=2t -\theta$,
$v=x^2 +y^2$, with $\theta=\arctan (y/x)$, and $u=U(y-t)$, $v=x$,
where $U$ is an arbitrary regular function.


\subsection{General remarks}

As a general remark  about the problem examined in this section
let us point out that the set of equations \eqref{EU} are of the
interest  for the study of the nonlinear dynamics of f\/luid plasma
conf\/igurations  and in particular for the interaction of
magnetized plasma vortices and for the development of
collisionless magnetic f\/ield line reconnection. In particular, the
explicit solutions found in the present article using Lie point
symmetries have a direct physical interpretation;  for instance,
the solutions (\ref{VW})--(\ref{Cor})  have a role in the recent
investigation of nonlinear evolution and saturation of magnetic
reconnection instabilities \cite{P3,P4} as plasma conf\/igurations
forced from the boundaries. We  note that  equations~\eqref{EU}
obey   boundary conditions that are determined by  values of the
electric and magnetic f\/ields at the boundaries of the domain where
the plasma is enclosed. These boundary conditions  may include
plasma f\/luxes through the boundaries. Referring for example to the
solutions (\ref{VW})--(\ref{Cor}),  we see that they correspond
to a  plasma velocity pattern controlled by the  boundary
conditions that determine the arbitrary functions~$A(t)$ and
$B(t)$.  The required boundary conditions correspond to imposing
the electric f\/ield  at these plates: the tangential  component  is
proportional to $\nabla v$  and determines the velocity pattern,
while  the normal component is proportional to $\partial
u/\partial t$ and determines the magnetic terms.

It is interesting to observe that the system \eqref{EU} can be
seen as a generalization of  the two-dimensional limit (i.e., the
case of solutions independent of the coordinate $z$ orthogonal to
the plane $(x,y)$),  of  the so-called Reduced Magnetohydrodynamic
equations (Kadomtsev--Pogutse equations) \cite{KPo}, see also
\cite{a15}. The symmetry properties of these equations, that are
frequently used to describe plasmas embedded in a very  strong
magnetic f\/ield, have been studied in  \cite{Sam}, see also
\cite[Vol.~2, p.~390]{Ib}. It can be remarked that a common
feature of the vector f\/ields describing the symmetry properties of
all these equations is the presence in their expression of one or
more arbitrary functions (see also~\cite{Bog}) constant along
magnetic (and velocity) f\/ield lines. The symmetry properties of
the three dimensional visco-resistive magnetohydrodynamic
equations in the incompressible limit have been studied
in~\cite{NuMHD}, see also  \cite[Vol.~1, p.~389]{Ib}.

Finally, the study of the partial symmetry (\ref{pp}) has been
useful not only to obtain a reduction of the initial equations,
leading also to new particular solutions, but also to point out a
special property of the system \eqref{EU}: it can be suitably
truncated as in equation \eqref{a1} (i.e.\ a physical parameter
can be put equal to zero) without altering its general symmetry
properties.

\section{Symmetry classif\/ication of an equation\\ for axisymmetric plasma
conf\/igurations}

Our third example deals with a completely dif\/ferent situation. The
equation we are going to examine contains indeed two arbitrary
functions $F(u)$ and $G(u)$ of the unknown variable $u=u(x,y)$,
and the goal is now to perform the symmetry classif\/ication of this
equation, i.e.\ to f\/ind those $F$, $G$ for which the equation
admits nontrivial symmetries. In general, the symmetry properties
of an equation may strongly depend on the choice of the arbitrary
functions involved.

Just to recall brief\/ly a signif\/icant example, let us point out the
case of the nonlinear Laplace equation $\nabla^2 u=F(u)$, with
$u=u(x,y)$ (or its hyperbolic counterpart, i.e.\ the nonlinear
wave equation $u_{xx}-u_{yy}=F(u)$). Symmetry properties of
equations of this form have been exensively studied; a complete
symmetry classif\/ication of the  nonlinear wave equations can be
found in~\cite{FuS}. Summarizing the results, one has that these
equations, excluding for a moment the particular cases $F=u$ and
$F=1$,  admit the inf\/inite dimensional algebra of symmetries
already mentioned (Section~2) if $F(u)=\exp(u)$ (the Liouville
case), the nearly trivial symmetry\[ X = (k-1)\left(
x\frac{\partial}{\partial x}+y\frac{\partial}{\partial y}\right) -2u
\frac{\partial}{\partial u}
\]
if $F(u)=u^k$, and only the obvious  symmetries (translations and
rotations of the variables~$x$, $y$) otherwise. In the case $F=u$
one f\/inds the standard symmetries of the linear equations;
f\/inally, if $F=1$, the inf\/inite-dimensional algebra of symmetries
given by \eqref{XL} is replaced by
\[
 X = \xi(x,y)\frac{\partial}{\partial
x}+\eta(x,y)\frac{\partial}{\partial y}
+\big(c\,u+A(x,y)\big)\frac{\partial}{\partial u},
 \]
where $c={\rm const}$  and $A(x,y)$ must satisfy
\[
\nabla^2A-2\xi_x+c = 0 .
\]

Performing the symmetry classif\/ication of a given equation may be
not an easy task: for some explicit examples and the presentation
of the techniques used in each situation, see
e.g.~\cite{NP,PY,PIv,Gu,ZL} and the papers quoted therein.

The PDE we want to consider is
\begin{gather} \label{FG}
u_{xx}+\frac{a}{x}u_x+u_{yy}=x^{2p}F(u)+G(u) ,
\end{gather}
where $a$, $p$ are constants. With $a=-1$, $p=1$ this equation is
known in plasma physics as the Grad--Schl\"uter--Shafranov
equation (see~\cite{Wes}) and describes  the  magnetohydrodynamic
force balance in a magnetically conf\/ined  toroidal plasma. In this
context, $u$ is the so-called magnetic f\/lux variable,  $x$ is a
radial  variable, then $x\ge 0$, while the two arbitrary f\/lux
functions $F(u)$, $G(u)$ are related to the plasma pressure and
current density prof\/iles.

According to the standard  def\/initions and procedure (see
\cite{Ov}), we can now look for the equi\-va\-len\-ce group;
preliminarily, we look for the kernel of the full groups of
equation~(\ref{FG}), i.e.\ the intersection of all groups admitted
by (\ref{FG}) for any arbitrary choice of $F$ and $G$. It turns
out that this kernel is almost trivial: it contains indeed only
the translation of the variable $y$.

The complete symmetry analysis of the above equation (\ref{FG}) is
a little bit tedious, but some simplif\/ication is provided  for
equations of this type, where the nonlinearity is present only in
the arbitrary functions $F$, $G$. In this case indeed one can show
that the coef\/f\/icients $\xi$, $\eta$, $\zeta$ of the admitted
symmetries $X=\xi\partial/\partial x+\eta\partial/\partial
y+\zeta\partial/\partial u$ are subjected to some precise linear
constraints, which admit a nice geometrical
interpretation~\cite{CXX}.

In the case of our equation (\ref{FG}), one obtains that, apart
from the translation of the variable~$y$, the equivalence group
contains the following transformations:

$i)$ the translation of the dependent variable $u$, i.e.\ $u\to
u+k$,

$ii)$ the scaling
\[ u\to\alpha u , \qquad F\to\alpha F , \qquad G\to\alpha G, \]

$iii)$ the scaling
\[ x\to\alpha x , \qquad y\to\alpha y , \qquad F\to\alpha^{-2(p+1)} F ,
\qquad G\to\alpha^{-2} G  .\] The conclusion can be summarized as
follows.
\begin{proposition}
Apart from the particular cases where both  $F$ and $G$ have the
form $c_1+c_2u$, to be examined later,  the only choices for the
functions $F$, $G$ which lead to an equation admitting some symmetry
(up to transition to equivalent functions via equivalence group)
are the following:
\[ \mbox{$a)$} \quad F(u) = u^{1+(p+1)/q } , \qquad G(u) = u^{1+1/q}\]
for all $q\not=0$, where the admitted symmetry is
\[
X=x\frac{\partial}{\partial x}+y\frac{\partial}{\partial y}-2qu
\frac{\partial}{\partial u} \] and
\[ \mbox{$b)$} \quad
F(u)=\exp \big( (1+p) u\big),\qquad G(u)=\exp (u)
\]
with symmetry
\[ X=x\frac{\partial}{\partial x}+y\frac{\partial}{\partial
y}-2\frac{\partial}{\partial u} . \]

In addition, one has that

$c)$ if in equation \eqref{FG} $p=-1$ and $G(u)\equiv 0$, then the
scaling of the variables $x$, $y$, generated by
\[ X = x{\partial\over{\partial x}}+y {\partial\over{\partial y}} \]
is a symmetry for all $F(u)$.
\end{proposition}

For instance, with account of the equivalence group
transformations, the most general choice of functions in class
$a)$, is
\begin{gather} \label{dil}
 a') \quad
F(u)=c_1\big(c+ u\big)^{1+(p+1)/q },\qquad G(u)=c_2\big(c+
u\big)^{1+(1/q)},
\end{gather}
where $c$, $c_1$, $c_2$ are constants, with symmetry\[
X=x\frac{\partial}{\partial x}+y\frac{\partial}{\partial
y}-2q(c+u) \frac{\partial}{\partial u} .\]

No new symmetry appears with particular choices of the parameters
$a$ and $p$, and of the functions $F$, $G$ as well, apart from the
following almost obvious exceptions.  First of all, we exclude
from our classif\/ication the case $a=0$ and $p=0$ (or $a=0$ and
$F(u)\equiv 0$) because in this case our equation coincides with
the already mentioned and well known nonlinear Laplace equation.
Other possibilities appear if both  $F$ and $G$ have the form
$c_1+c_2u$. If $F=G=u$, the equation exhibits the standard
symmetries of any linear equation. If $F=1$, $G=u$, there is a~new
admitted symmetry which can be written in the following way:
\begin{gather} \label{u1}
X = \big( u+\Phi(x,y)\big)\frac{\partial}{\partial u},
\end{gather} where, with ${\cal E}(u)$ denoting the left hand side
of equation \eqref{FG}, $\Phi$ solves the equation\[ {\cal
E}(\Phi)=\Phi-x^{2p}  .\] If $F=u,\ G=1$, the new symmetry has the
same form as \eqref{u1}, but $\Phi$ must solve
\[ {\cal E}(\Phi)=x^{2p}\Phi-1 .\]
Thanks to the equivalence group of our equation, it remains only
to consider the case \mbox{$F=G=1$}. In this case, although of no
great interest in the applications, the situation is much more
complicated. Indeed, the most general vector f\/ield $X$ belonging
to the algebra of the admitted symmetries can be globally
described in the following way:
\[ X = (c_1x+2c_2xy)\frac{\partial}{\partial x}
+\big(c_0+c_1y-c_2\big(x^2-y^2\big)\big) \frac{\partial}{\partial
y} +\big((c_3-c_2ay)u+\Psi(x,y)\big)\frac{\partial}{\partial u},
\] where the function $\Psi(x,y)$ satisf\/ies the equation\[ {\cal
E}(\Psi)+c_3-2c_1+
x^{2p}\big(c_3-2c_1(p+1)\big)-c_2y(4+a)-c_2x^{2p}y\big(4(p+1)+
a\big) = 0   . \]

There is, in addition, an interesting conditional symmetry for
equation (\ref{FG}) in the case $a=-p\not=0$  (which covers in
particular the concrete case $a=-p=-1$ of the
Grad--Schl\"uter--Shafranov equation), and $G(u)=k^2F(u)$ ($k={\rm
const}$). The conditional symmetry is given~by
\[ X= k\frac{\partial}{\partial x}+x^p\frac{\partial}{\partial y}  . \]
This symmetry is particularly interesting and useful because, with
introduction of  the symmetry-invariant variable
\[ s= \frac{1}{p+1}x^{p+1}-ky \qquad   {\rm if} \quad p\not=-1,
\quad {\rm or} \quad  s=\ln x-ky    \quad  {\rm if}\ p=-1 \] the
PDE is transformed into the very simple reduced ODE
\[ u_{ss}=F(u) \]
which can be easily integrated.

The class $(a)$  (or $(a')$) symmetry in Proposition 4
corresponds, for $a=-p=-1$, i.e.\  for the
Grad--Schl\"uter--Shafranov equation, to spatial dilations  where
$u$ scales as $x^{-2q}$, $y^{-2q}$. An elementary solution
obeying such a symmetry is obtained  by taking, e.g., $q=-2$ and
by considering ``cylindrical'' solutions  with $\partial
u/\partial y = 0$. Choosing  $F(u)=c_1 $ and $G(u)=(8-c_1)
u^{1/2}$, according to equations~(\ref{dil}), we obtain
from~\eqref{FG} $u= x^4$. By recalling the physical meaning of the
functions $F(u)=c_1 $ and $G(u)=(8-c_1) u^{1/2}$ we f\/ind that this
solution  describes  a plasma conf\/iguration in the (dilation
dependent) domain $0 < x < x_0$ with a pressure $p$ of the form $p
= (c_1/ 4\pi) (x_0^4 - x^4)$ and a total current $I$  along the
cylinder of the form $ I^2 = [4 (8-c_1)/3] (x_0^6 - x^6)$, where
$[4 (8-c_1)/3]^{1/2} x_0^3$ is the current f\/lowing along  a
conductor  at $x=0$.  The boundary conditions have been chosen
such that the pressure and the total current vanish at $x= x_0$.

Finally we remark that the Grad--Schl\"uter--Shafranov equation is
a special example of a~family of equations that are used in plasma
physics and in space physics and astrophysics~\cite{Lovelace} in
order to describe time independent conf\/igurations of a plasma
embedded in a magnetic f\/ield in the presence of plasma f\/luid
motions (not included in the Grad--Schl\"uter--Shafranov
equation).  Similarly to the Grad--Schl\"uter--Shafranov equation,
these equations depend on a set of arbitrary functions, but can
change from elliptic  to hyperbolic type depending on the (local)
value of the plasma velocity.

\section{Conclusion}

In conclusion, we believe that the three examples considered in
this paper may give a detailed idea of the opportunities of\/fered
by symmetry methods to the study of dif\/ferential equations arising
from modern physical theories. Let us  emphasize in particular the
completely dif\/ferent peculiarities of our examples, all suggested
by plasma physics, and correspondingly the dif\/ferent role played
by symmetry techniques in their application to each situation. For
other dif\/ferent applications to plasma physics of similar methods
and procedures,  based on general symmetry properties, we refer
also, e.g., to the papers \cite{Mul,KKP,K3,K2}.

\subsection*{Acknowledgements}
We are indebted to the referees for their useful comments and
valuable suggestions. We also thank Nino Valenti for his precious
help with a search in the literature.

\LastPageEnding

\end{document}